# Control principles of metabolic networks

Georg Basler[1,2*], Zoran Nikoloski[3*], Abdelhalim Larhlimi[4], Albert-László Barabási[5-7], Yang-Yu Liu[6-7]

[1] Department of Environmental Protection, Estación Experimental del Zaidín CSIC, Granada, Spain.
[2] Department of Chemical and Biomolecular Engineering, University of California, Berkeley, California, USA.
[3] Systems Biology and Mathematical Modeling, Max Planck Institute of Molecular Plant Physiology, Potsdam, Germany.
[4] Laboratoire d'Informatique de Nantes Atlantique, Université de Nantes, France.
[5] Center for Complex Network Research and Departments of Physics, Computer Science and Biology, Northeastern University, Boston, Massachusetts, USA.
[6] Channing Division of Network Medicine, Department of Medicine, Brigham and Women's Hospital, Harvard Medical School, Boston, Massachusetts, USA.
[7] Center for Cancer Systems Biology, Dana-Farber Cancer Institute, Harvard Medical School, Boston, Massachusetts, USA.

*These authors contributed equally to this work.

Correspondence and requests for materials should be addressed to A.-L. B. (email: alb@neu.edu) or Y.-Y.L. (email: yyl@channing.harvard.edu).

## Abstract

Deciphering the control principles of metabolism and its interaction with other cellular functions is central to biomedicine and biotechnology. Yet, understanding the efficient control of metabolic fluxes remains elusive for large-scale metabolic networks. Existing methods either require specifying a cellular objective or are limited to small networks due to computational complexity. Here we develop an efficient computational framework for flux control by introducing a complete set of flux coupling relations. We analyze 23 metabolic networks from all kingdoms of life, and identify the driver reactions facilitating their control on a large scale. We find that most unicellular organisms require less extensive control than multicellular organisms. The identified driver reactions are under strong transcriptional regulation in *Escherichia coli*. In human cancer cells driver reactions play pivotal roles in tumor development, representing potential therapeutic targets. The proposed framework helps us unravel the regulatory principles of complex diseases and design novel engineering strategies at the interface of gene regulation, signaling, and metabolism.

## Introduction

Understanding how cellular systems are controlled on a genome scale is a central issue in biology and medicine. Metabolic networks are at the center of systems biology approaches to unravel cellular control because metabolism carries the life-sustaining cellular functions shaping the molecular phenotype[1]. The steady-state principle and physico-chemical constraints (e.g. mass balance and thermodynamics) have been employed to reduce the number of considered network states, facilitating the prediction of genotype-phenotype relationships and intervention strategies for biotechnological or medical purposes[2]. In particular, flux balance analysis and variations thereof have been successfully applied to the metabolic networks of unicellular organisms to predict their metabolic and cellular phenotypes[3]. Yet, those approaches are *biased*[4], because they restrict the flux space to an *a priori* specified reference state by assuming a cellular objective to be optimized by the organism[5]. While optimization of biomass yield has proven useful for unicellular organisms, identification of a suitable objective for multicellular organisms remains a non-trivial endeavor[1]. Other approaches, e.g. elementary flux modes[6] and extreme pathways analyses[7], do not assume a cellular objective and hence are *unbiased*. However, despite extensive studies and recent advances[8], these unbiased approaches are limited to rather small networks due to their computational complexity. It remains unclear how complex biological systems efficiently coordinate metabolism on a large scale.

Here we develop a framework to systematically study the control principles of large-scale metabolic networks. The key idea is that the activity of a reaction can be controlled by directly manipulating a reaction to which it is coupled[9]. The most efficient control strategy is then given by the smallest set of *driver reactions* that must be directly controlled for controlling the activity of all reactions in the network at steady state. To reveal the smallest set of driver reactions offering control over the whole network, we first need to fully exploit all possible qualitative couplings among reactions. There are four possible cases by which the flux of one reaction $R_1$ can be used to qualitatively control the flux of another reaction $R_2$: (i) an active flux of $R_1$ can activate $R_2$; (ii) an inactive flux of $R_1$ can deactivate $R_2$; (iii) an inactive flux of $R_1$ can activate $R_2$; and (iv) an active flux of $R_1$ can deactivate $R_2$. We find that the flux coupling types proposed and widely used in literature only account for cases (i) and (ii), unaware of the potential offered by cases (iii) and (iv). Here we identify two new coupling types that describe well-known biochemical principles and allow us to consider the remaining two cases, eventually obtaining a *complete* set of qualitative coupling relations. We show that the driver reactions can be determined efficiently even for metabolic networks involving thousands of reactions by solving a classical graph-theoretic problem via integer linear programming. Our framework does not require any *a priori* knowledge of the cellular objectives, and hence is *unbiased*. It is also computationally efficient, facilitating, for the first time, systematic analyses of the control principles of large-scale metabolic networks and offering mechanistic insights into cellular regulation.

## Results

**A complete set of flux coupling relations enables efficient control of metabolic networks.** Formally, the structure of a metabolic network is uniquely specified by its $m \times n$ stoichiometric matrix, $S=[s_{ij}]$, with $m$ rows denoting metabolites and $n$ columns representing reactions. An entry $s_{ij}$ represents the molarity of metabolite $i$ in reaction $j$, with negative entries denoting substrates and positive entries indicating products. A feasible flux distribution of a metabolic network $S$ is defined as a flux vector $v$ satisfying the steady-state condition ($Sv=0$) subject to lower and upper bounds ($lb \leq v \leq ub$). We propose an additional constraint of a non-zero exchange of matter with the environment (i.e. the flux $v_i \neq 0$ for at least one exchange reaction $i$). This is a natural requirement for any living system, which also allows us to introduce new flux coupling relations (see *Methods*). The *status* $\sigma_i^v=|\text{sign}(v_i)|$ of a reaction $i$ in $v$ is called *active*, if $\sigma_i^v=1$, and *inactive*, if $\sigma_i^v=0$. The aforementioned constraints imply that some reactions must operate in a concerted manner at steady state, leading to coupling relations between rates, and, thus, status of reactions.

To represent the coupling relations between reactions in a metabolic network, we construct the flux coupling graph (FCG)[9], where vertices denote reactions and edges describe the coupling types (Fig. 1a and *Methods*). Three types of flux coupling have been proposed in literature[9]: directional, partial, and full coupling. A reaction $i$ is *directionally coupled* to $j$ if $\sigma_i^v=1$ implies that $\sigma_j^v=1$ (and equivalently, $\sigma_j^v=0$ implies $\sigma_i^v=0$) (e.g. $R_1$ and $R_3$ in Fig. 1a). Partial coupling is a special case of directional coupling: Two reactions, *i* and *j*, are *partially coupled* if they have the same status, i.e. $\sigma_i^v=\sigma_j^v$, in every feasible flux distribution (e.g. $R_2$ and $R_3$ in Fig. 1a). Moreover, full coupling is a special case of partial coupling: two reactions are *fully coupled*, if there is a constant, $\lambda\neq 0$, such that $v_i=\lambda v_j$ for every feasible flux distribution $v$ (e.g. $R_4$ and $R_5$ in Fig. 1a). Thus, full and partial coupling have equivalent implications with respect to the status of reactions *i* and *j*, since in both cases $\sigma_i^v=1$ if and only if $\sigma_j^v=1$. Moreover, these three coupling relations are similar in the sense that they allow a reaction to be activated or deactivated by imposing the same status on a reaction to which it is coupled ($\sigma_i^v=\sigma_j^v$). But they do not allow for activating or deactivating a reaction by using reactions that have a different status ($\sigma_i^v\neq\sigma_j^v$). Hence, they account only for two of the four possible cases of qualitative coupling, and consequently, do not provide a complete set of qualitative coupling relations (see *Flux coupling analysis* in Supplementary Methods).

To consider the two remaining cases, we introduce two new coupling types, called anti- and inhibitive couplings: Two reactions *i* and *j* are *anti-coupled*, if at least one of *i* and *j* is active in any feasible flux distribution (i.e., $\sigma_i^v=0$ implies $\sigma_j^v=1$ and $\sigma_j^v=0$ implies $\sigma_i^v=1$, cf. $R_3$ and $R_5$ in Fig. 1a). A reaction *i* is *inhibitively coupled* to a reaction *j* if a maximum flux of reaction *i* implies that *j* is inactive. Note that just activating a reaction cannot imply the deactivation of another reaction (see *Flux coupling analysis* in Supplementary Methods). Inhibitive coupling accounts for the competition of coupled reactions for common substrates (e.g. $R_4$ and $R_1$ in Fig. 1a) or the accumulation of common products (e.g. $R_2$ and $R_1$ in Fig. 1a). The two new coupling relations allow a reaction *i* to be activated or deactivated by using a reaction *j* which has a different status ($\sigma_i^v\neq\sigma_j^v$). Since the resulting five coupling types now account for each of the four possible cases of qualitative coupling between pairs of active or inactive reactions they offer a *complete* set of qualitative coupling relations.

The coupling relations imply that reaction fluxes may not only be directly controlled by the regulation of enzyme activities and concentrations[10], but also be implicitly controlled by the requirement of achieving and maintaining a steady state. This suggests that the expression of genes encoding the enzymes of coupled reactions is coordinated in agreement with their imposed status. For example, the synthesis of the enzyme catalyzing a reaction which is inactive due to its coupling with other reactions would imply a waste of resources. Likewise, if a reaction is required to be active due to the present couplings, but cannot occur based on the existing regulatory interactions, then the system cannot operate at steady state. Thus, we hypothesize that coupled reactions are co-regulated by a common transcription factor (TF). To test this hypothesis, we overlay the gene regulatory network of *E. coli*[11] on its genome-scale metabolic network[12] and determine the agreement between reaction couplings and the co-regulation of their enzyme-coding genes by a common TF (see *Methods*). This generalizes a previous study of the operonic organization of genes associated with fully and directionally coupled reactions[13], since TFs may regulate genes across operons. We find that coupled reactions are very likely co-regulated (hypergeometric test, p-value=$3.3 \cdot 10^{-290}$, including all types of coupling). With respect to the individual coupling types, there is a striking agreement between co-regulation and fully coupled reactions (Supplementary Fig. 2): while only 24,303 (5.7%) of the total 427,350 pairs of enzyme-catalyzed reactions are co-regulated, 175 (42.9%) of the 408 fully coupled pairs are co-regulated, which is highly significant (hypergeometric test, p-value=$4.2 \cdot 10^{-105}$). In contrast, none of the 36 partially coupled reaction pairs is co-regulated. There is a small, but significant (p-value=$2.6 \cdot 10^{-2}$) agreement for the 2,999 directionally coupled reactions, of which 196 are co-regulated (6.5%). This extends the finding that genes associated with fully and directionally coupled reactions are frequently located on the same operon[13] and suggests their coordinated regulation by common TFs. While there are no anti-coupled reactions in the genome-scale metabolic network of *E. coli*, the proposed inhibitive coupling also shows a strong agreement with co-regulation: of the 6,847 inhibitively coupled reaction pairs, 1,220 are co-regulated (17.8%, p-value=$7.9 \cdot 10^{-278}$). These results demonstrate that the expression of metabolic genes is coordinated by transcriptional regulation in line with reaction couplings, confirming their relevance for metabolic regulation.

**Flux coupling profiles capture phylogenetic signals of metabolic networks.** To determine if the *complete* set of qualitative coupling relations reflects functional principles of metabolism, we analyze 23 high-quality metabolic networks from all kingdoms of life (Table 1). The networks are selected based on their model quality, curation, and expert knowledge, and vary greatly in their sizes, tissues, subcellular compartmentalization, and uni- or multicellular organization. For each network, we first calculate the five coupling types over all reaction pairs. We find that anti-coupling is rarely found in genome-scale networks (Supplementary Table 1), which indicates the existence of redundant pathways operating in steady states of these networks (see Supplementary Methods). The relative frequencies of the five coupling types in a given network define the corresponding *flux coupling profile* (see *Methods*).

To determine whether the flux coupling profiles reflect functionally important features of the analyzed metabolic networks, we assess their statistical significance using network randomization under mass-balance constraints[14]. We find that the flux coupling profiles for most networks and coupling types are significantly different from their randomized variants, suggesting that they reflect evolutionary features of their metabolism[15] (see *Methods*). To determine the extent to which flux coupling profiles reflect the phylogeny of the considered biological systems, we apply three classical clustering algorithms together with three well-established cluster quality measures (see *Methods*). The clustering of the flux coupling profiles indeed reveals the established phylogeny and functional differences of the metabolic networks, as demonstrated by the separation into five clusters obtained consistently from the three clustering algorithms (Fig. 2): the two flux coupling profiles of non-extremophile archaea (cluster 1) are separated from those of extremophiles (cluster 2) and the two networks of *E. coli* central metabolism (cluster 3). The flux coupling profiles of three cancer types fall together with the profiles of the genome-scale metabolic networks of *E. coli* and *M. tuberculosis* (cluster 4). Moreover, the cancer networks (clusters 2 and 4) are clearly separated from the corresponding healthy tissues (cluster 5), suggesting that distinct coupling patterns characterize healthy and cancer tissues. In particular, the flux coupling profiles of all 9 considered eukaryotes form a separate branch of the clustering tree (cluster 5) despite the apparent differences in the size and cellular organization of the represented metabolic networks (cf. Table 1).

The frequencies of different types of coupling thus allow us to distinguish functional features of metabolism (Fig. 2b-f): non-extremophile archaea (cluster 1) show a balance of full, partial, directional, and inhibitive couplings, while the metabolism of extremophiles (cluster 2) is dominated by directional couplings. The prevalence of full and inhibitive couplings in cluster 3 suggests a balance between tightly coupled and competing pathways in *E. coli* central metabolism. This is in contrast to its genome-scale metabolic network (cluster 4), suggesting the prevalence of distinct coupling patterns in primary and secondary metabolism. Moreover, cluster 5 is dominated by inhibitive couplings, which suggests that competing pathways, a widespread feature of eukaryote metabolism due to gene and whole genome duplications[16], are key to controlling these networks. By contrast, the lower frequency of inhibitive couplings in the metabolic networks of cancer cells (clusters 2 and 4) allows a clear distinction from the healthy tissue networks.

For comparison, we find that the clustering based only on the previously studied flux coupling types, i.e. full, partial, and directional couplings, does not correspond to phylogenetic or functional relationships of the analyzed metabolic networks (see *Methods* and Supplementary Fig. 3). Here, we observe that networks within functionally or phylogenetically related groups, such as the networks of *E. coli* central metabolism, extremophiles, or eukaryotes, are dispersed across clusters. Moreover, the separation of cancer networks from their corresponding healthy tissue networks is no longer apparent, since cancer networks are assigned to clusters together with the corresponding healthy tissue or other eukaryotic networks. This indicates that the flux coupling profiles obtained from the complete set of coupling types, but not those of the previously studied coupling types, reflect an intrinsic phylogenetic signal of metabolic networks. Finally, we find that the clustering based only on structural determinants (i.e. of the cumulative distribution of singular values[17,18]), yields few large clusters that are significantly different from those of the complete set of flux coupling profiles (adjusted Rand Index of -0.06, see *Methods*) and, too, lack correspondence to phylogenetic relationships. Hence, a characterization of the phylogenetic or functional requirements of these networks based only on their structural determinants is not obvious.

**Driver reactions are obtained by integrating flux coupling with activity patterns.** Based on the complete set of qualitative coupling relations, we propose a comprehensive control framework to identify the smallest sets of driver reactions. Consider a *reaction activity pattern* $\sigma=\{0,1\}^n$ that describes the status of reactions in a given metabolic network at steady state (Fig. 1b-d). We can generate feasible activity patterns using two distinct sampling schemes (see *Methods*). For each feasible activity pattern, we determine its driver reactions as follows. First, we construct a *control graph* (CG) by integrating the FCG with the activity pattern: The CG contains a vertex for each reaction in the metabolic network and a directed edge ($i \rightarrow j$), if controlling the status of reaction $i$ allows us to impose the status of reaction $j$ in the activity pattern. For example, if reaction $i$ is directionally coupled to reaction $j$, and $\sigma_i=\sigma_j=1$ according to the desired activity pattern, then the CG contains the directed edge $(i, j)$, but not $(j, i)$, as $i$ can be used to activate $j$, but not necessarily vice versa (e.g. $R_1$ and $R_2$ in Fig. 1b, see *Methods*). Then a smallest set of driver reactions, denoted as $D(CG)$, which must be directly controlled to impose the status of all reactions in the activity pattern, is just a smallest vertex set $D$ in CG such that the union of $D$ and its out-neighborhood $N^+(D)$ covers all vertices in CG, i.e.

$$D(CG) = \min_{|D|}\{D \subseteq V(CG): D \cup N^+(D) = V(CG)\}.$$

In graph theory, $D(CG)$ is called a *minimum out-dominating set*[19] of the CG, which can be exactly solved via an integer linear program (see *Methods*). Although in general the minimum out-dominating set problem is NP-hard[20], we find that it is actually computationally inexpensive even for large networks with thousands of vertices (see *Efficient calculation of driver reactions* in Supplementary Discussion).

To quantify the difficulty of qualitative flux control in metabolic networks, we calculate the average size of the smallest driver reaction sets over 1,000 feasible activity patterns. Inspection of the relationship between network size and the average number of driver reactions over different activity patterns indicates that the complexity of qualitative flux control scales linearly with the size of metabolic networks (Pearson correlation of 0.97, p-value=$3 \cdot 10^{-14}$, Supplementary Fig. 4). However, when considering the fraction of driver reactions, we observe that the genome-scale metabolic networks of most prokaryotes require smaller fractions of drivers (15.5-43.4%), and, thus, are easier to control, in comparison to those of eukaryotes (35.8-49.2%). Moreover, it is more facile to control the considered genome-scale metabolic networks than the smaller networks of central metabolism (15.5-49.2% and 57.9-82.6%, respectively, see Fig. 3a). This finding is in line with the observation that, unlike primary metabolism, the pathways of secondary metabolism may be controlled by manipulating one or few enzymes[1].

In addition, we observe that the fraction of driver reactions follows a power-law decay with the fraction of coupled reactions (Fig. 3b, scaling coefficient of 0.24, Pearson correlation of –0.72 (log-log), p-value=$1.2 \cdot 10^{-4}$). On one hand, this finding suggests that structural modifications leading to few additional coupled reactions may substantially improve the control of metabolic networks that are otherwise difficult to control. On the other hand, this also indicates that the inclusion of additional coupled reactions has only a small impact on networks that have few driver reactions. These results are robust with respect to the choice of the sampling scheme and the number of sampled activity patterns (see *Methods*). Despite the finding that flux coupling analysis is sensitive to missing reactions[21], we observe that the determined driver reactions are not significantly altered (Pearson correlation of driver frequencies, see *Sensitivity to missing reactions* in Supplementary Discussion) when removing up to 15% of the reactions from the network of *E. coli* central metabolism, or up to 25% of the reactions from the genome-scale metabolic network. Altogether, these robustness tests demonstrate the power of our method in making predictions from incomplete network reconstructions[22].

**Driver reactions are under strong transcriptional regulation.** Since the key idea of our method is that driver reactions facilitate control of metabolic fluxes, we hypothesize that the enzyme-coding genes associated with driver reactions are under strong transcriptional regulation to achieve efficient control of metabolism. To test this hypothesis, we determine the number of TFs regulating the enzyme(s) associated with each reaction of the genome-scale metabolic network of *E. coli*. Similar to the classification of vertices by their roles in controlling an arbitrary complex network[23], we group the reactions in the metabolic network of *E. coli* into three classes depending on the number of activity patterns in which they appear as driver reactions: *redundant reactions* are never required for control and, thus, can always be controlled through other reactions; *intermittent reactions* are driver reactions for at least one, but not all considered activity patterns, and *critical reactions* are driver reactions for all activity patterns. We find that both critical and intermittent reactions are regulated by significantly more TFs (average of 1.9 and 2.9, respectively) in comparison to redundant reactions (average of 1.0, Wilcoxon rank sum test p-values of $1.3 \cdot 10^{-15}$ and $2.1 \cdot 10^{-37}$, respectively) (Fig. 4a). The large number of TFs associated with intermittent reactions indicates that reactions whose role as driver or non-driver varies for different activity patterns require the most extensive transcriptional regulation. In addition, we observe that fewer TFs than expected by chance (46 of 82) are associated with more than one class of reactions (Fig. 4b, permutation test, p-value=$10^{-3}$), suggesting that the three classes of reactions operate under distinct modes of transcriptional regulation. Moreover, of the TFs associated with more than one class, significantly more are shared between intermittent and critical reactions (34 of 46) compared to the other intersections (30 TFs are shared between redundant and intermittent, and 24 between redundant and critical reactions; permutation test, p-value=$2 \cdot 10^{-3}$), which points at the similarly important role of intermittent and critical driver reactions in transcriptional regulation, and attributes a less pronounced role to redundant reactions.

**Driver reactions control tumor development at the interface between gene regulation, signaling, and metabolism.** The accelerated glucose uptake, glycolysis, and lipogenesis are a signature of proliferating cells in a variety of tumors, thought to arise through aberrant interactions between gene regulation, signaling pathways, and metabolism[24]. Indeed, we find that the four investigated networks of cancer metabolism have 65% (1,799/2,778) of reactions

in common (Supplementary Fig. 5a), suggesting the involvement of similar metabolic pathways. However, only 40% (300/742) of the critical driver reactions are shared among the four cancer types (Supplementary Fig. 5b). The discrepancy between the identified driver reactions is most pronounced for the lung and urothelial cancer models, which share 81% of reactions, but only 58% of their driver reactions. This suggests that the driver reactions may reveal control principles of different types of cancer beyond their universal metabolic requirements.

To determine if the identified driver reactions are related to genes known to cause cancer[25], we rely on the gene-enzyme-reaction associations. Remarkably, we find that critical driver reactions in the four cancer networks are frequently associated to cancer causing genes (hypergeometric test, p-value=$6.8 \cdot 10^{-4}$), suggesting that our framework can be used to identify reactions whose causal role in cancer is hitherto unknown. In contrast to previous analyses that compare the presence and absence of reactions in cancer vs. healthy tissues[26], here we pinpoint reactions that are involved in the development of cancer. We obtain similar results when extending the analysis to include cancer associated genes from high-throughput mutational screenings, whole-exome and whole-genome sequencing[27] (hypergeometric test, p-value=$9.6 \cdot 10^{-4}$, see *Methods*).

We hypothesize that those reactions that are critical drivers in networks of cancer metabolism but are redundant in the corresponding healthy tissue play key roles in tumor development. We refer to these reactions as *metabolic switches* because of their divergent role in controlling cancer, but not healthy metabolism. In total, there are 21 metabolic switches (Supplementary Table 2). The number of couplings of metabolic switches differs between the cancer and healthy tissue networks, mostly with respect to inhibitive couplings (Supplementary Fig. 6). Specifically, the lower frequency of inhibitive couplings in cancer compared to eukaryote metabolism (cf. Fig. 2) is also observed in metabolic switches. This suggests that the less extensive competition between pathways in cancer cells may be an important feature of their (de)regulation.

We focus on metabolic switches across cancer types, i.e. reactions which are critical drivers in more than one cancer type and redundant in the corresponding healthy tissues. In total, we find five such reactions (Supplementary Table 2): (1) hydrolysis of S-adenosylhomocysteine (SAH) by SAH hydrolase (EC 3.3.1.1); (2) phosphorylation of deoxyguanosine by deoxyguanosine kinase (dGK, EC 2.7.1.113); (3) phosphorylation of diacylglycerol (DAG) to phosphatidic acid (PA) by diacylglycerol kinase (DAGK, EC 2.7.1.107); (4) oxygenation of inosine monophosphate (IMP) by IMP dehydrogenase (EC 1.1.1.205); (5) phosphorylation of deoxyadenosine by deoxyadenosine kinase (dAK, EC 2.7.1.76). Indeed, four of these reactions are widely known to be involved in key aspects of cancer regulation. First, SAH hydrolase is essential for maintaining low levels of SAH in healthy cells, as varying levels of SAH can lead to DNA hypomethylation[28,29] and cancer[30-34]. Its inhibition by different drugs selectively induces apoptosis in cancer, but not in healthy cells[35-39], confirming its predicted role as critical driver in cancer, but not in healthy tissues. Second, high level of dGK is associated with leukemia[40], and its phosphorylation of several nucleoside analogues increases sensitivity to these anticancer drugs[41-43]. Third, DAGKs are enzymes catalyzing a key step of the phosphatidylinositol cycle, acting as a molecular switch between cell signaling and lipid metabolism[44]. Their activation is essential for invasiveness, mitogenesis and growth of various cancer types[45-47]. Elevated levels of DAGKs are associated with malignant transformation[44], and the product PA activates the cancer associated mTOR signaling pathway[48]. Fourth, IMP dehydrogenase has a central role in developing cancers from adult stem cells[49] and is widely used as therapeutic target for a range of cancer types[50-52]. Importantly, each of these enzymes has multiple roles in cancer signaling, gene regulation, and metabolism[36,41,48,51], suggesting that metabolic switches play key roles in regulation of cancer metabolism at the interface with other cellular functions. While the specific role of dAK in cancer is less evident, its substrate deoxyadenosine was found to be involved in leukemia[53,54], and the same reaction is also catalyzed by dGK and deoxycytidine kinase, which are known to be involved in tumor development[40,55]. It is also tightly linked to SAH metabolism and DNA methylation[56]. Our results therefore indicate that deoxyadenosine phosphorylation is also central to tumor development and may be used as a novel therapeutic target.

Importantly, our method also helps us reveal the molecular causes for the (de)regulation of metabolic pathways. For example, we find that SAH hydrolase is fully coupled to methyltransferase in the healthy glandular breast and urothelial networks, indicating that the production of SAH, a byproduct of essential methylation reactions, is coupled to its degradation by SAH hydrolase (Fig. 5a, b). Indeed, the tight coupling of SAH production by methylation reactions to its degradation by SAH hydrolase is necessary to avoid the inhibition of essential methyltransferases, and thus hypomethylation, because SAH strongly inhibits methyltransferases[28,29]. In the metabolic networks of the corresponding cancer types (i.e. breast and urothelial cancer), we find that the SAH hydrolase and methyltransferase reactions are not fully but inhibitively coupled due to the common production of SAH (since SAH hydrolysis is reversible) (Fig. 5c). This result suggests that SAH levels are no longer controlled by SAH hydrolase activity in these cancers. Instead, the inhibitive coupling from the reversed SAH hydrolase reaction to methyltransferase indicates that a large flux toward the production of SAH may inhibit methylation in cancer, which is in agreement with the experimental studies[28,34].

## Methods

**Analyzed metabolic networks.** We analyze 23 metabolic networks encompassing a broad spectrum of organisms from all kingdoms of life, including genome-scale networks and subsystems, condition- and tissue-specific, healthy, and cancer networks. The covered organisms include three archaea[57-59], five prokaryotes[12,60-63], and five eukaryotes[17,64-67].

The represented subsystems include three networks of central metabolism[5,66,68], four human tissue-specific networks and four cancer networks associated to these tissues[69]. The number of subcellular compartments ranges from one (in archaea, central metabolism of *E. coli*, *M. pneumoniae*, and *M. tuberculosis*) to seven in the *H. sapiens* genome-scale, tissue-specific and cancer networks. The number of reactions capable of carrying steady-state flux in the analyzed networks ranges from 75 (for central metabolism of *E. coli*) to 2,719 (for glandular breast tissue). Details of the analyzed networks are provided in Table 1.

**Calculation of flux coupling relations.** Full, partial, and directional coupling are calculated using the F2C2 tool[70]. To define anti-coupling, we restrict the set of feasible flux distributions to

$$F = \{v \in \mathbb{R}^n | Sv = 0, lb \leq v \leq ub, \exists i \in E: v_i \neq 0\},$$

where *n* is the number of reactions, *S* the stoichiometric matrix, *lb*, *ub* the lower and upper flux bounds, and *E* the set of exchange reactions. Here, we use *ub*=1,000, *lb*=0 for irreversible, and *lb*=-1,000 for reversible reactions, since these values haven previously been used for similar constraints[71]. We point out that reaction couplings do not depend on the choice of (uniform) flux bounds (Supplementary Methods). To determine whether two reactions *i* and *j* are anti-coupled, i.e., $v_i$=0 implies $v_j \neq 0$ for each feasible flux distribution *v*, we determine the feasibility of a vector $v \in F$ satisfying $v_i$=0 and $v_j$=0 using a mixed integer linear program (MILP). If no such vector exists, then the reactions *i* and *j* are anti-coupled. To determine whether a reaction *i* is inhibitively coupled to a reaction *j*, i.e., a maximum flux of reaction $i$ implies $v_j$=0, we first determine the maximum flux of $i$, $i^*=\max_{v \in F} v_i$, using flux variability analysis[72]. We then determine the feasibility of a vector $v \in F$ satisfying $v_i = i^*$ and $v_j \neq 0$ using a MILP. If no such vector exists, then $i$ is inhibitively coupled to $j$. The MILPs for calculating full, partial, directional, anti- and inhibitive couplings are discussed in the Supplementary Methods.

**Clustering of flux coupling profiles.** The flux coupling profile of a metabolic network $S$ is given by a vector $\phi \in \mathbb{R}^5$ representing the normalized frequencies of the five coupling types:

$$\phi_i(S) = \frac{|w_i|}{\sum_{j=1}^{5}|w_j|}, i = 1, \dots, 5,$$

where $|w_i|$ is the number of reaction pairs coupled by type $i$. We apply three classical clustering algorithms, i.e., hierarchical (agglomerative), k-means, and k-medoids, based on the Euclidean distances of the flux coupling profiles from the 23 analyzed networks. We evaluate the clusters obtained at each cutoff distance (hierarchical clustering) and for each possible number of clusters (k-means and k-medoids) using the Silhouette, Calinski-Harbasz[73], and Davies-Bouldin[74] indices. For a given clustering $C=\{C(1), ..., C(k)\}$ of our $n$=23 networks, the Silhouette index is given by

$$SI(C) = \frac{1}{n}\sum_{i=1}^{n}\frac{bcd_i - wcd_i}{\max\{bcd_i, wcd_i\}},$$

where $bcd_i$ is the smallest average of the Euclidean distances of a profile $i$ to the profiles in all other clusters, and $wcd_i$ is the average Euclidean distance of profile $i$ to all profiles within the same cluster. Thus, $SI(C) \in [-1,1]$, with larger values indicating higher quality of the clustering. The Calinski-Harbasz index[73] is defined as

$$CH(C) = \frac{bcv(C)}{wcv(C)} \cdot \frac{n-k}{k-1},$$

where $bcv(C)$ is the between-cluster variance, and $wcv(C)$ is the within-cluster variance. Thus, $CH(C) \geq 0$, where larger values indicate higher quality of the clustering. The Davies-Bouldin index[74] is given by

$$DB(C) = \frac{1}{k}\sum_{i \in C} \max_{j \neq i} \left\{\frac{wcc_i + wcc_j}{cd_{i,j}}\right\},$$

where $wcc_i$ is the average Euclidean distance between each profile in cluster $i$ and the centroid of cluster $i$, $wcc_j$ is the average distance of each profile in cluster $i$ and the centroid of cluster $j$, and $cd_{i,j}$ is the distance between the centroids of clusters $i$ and $j$. Thus, $DB(C) \geq 0$, where small values indicate a high quality of the clustering.

When applying the three clustering algorithms to the flux coupling profiles obtained from the complete set of coupling types, we obtain the same clustering from each algorithm only for $k$=5 clusters, which corresponds to a cutoff distance of 0.25. The clustering and quality for all other cutoffs differ depending on the selected algorithm and quality measure. For $k$=5, we obtain $SI(C)$=0.66, $CH(C)$=30.0, and $DB(C)$=0.7 (cf. Fig. 2a). The highest cluster qualities are obtained for $k$=20 (using the Calinski-Harbasz index with k-medoids or k-means clustering) and $k$=22 (all other combinations), corresponding to only 1 or 3 non-singleton clusters formed by 2 networks each. This indicates a strong separation of most networks by their flux coupling profiles. The non-singleton cluster for $k$=22 is formed by the two non-extremophile archaea, while $k$=20 yields two additional clusters formed by the healthy glandular breast and lung tissue networks, and the healthy urothelial and kidney tissue networks, respectively. Each of these clusters is a subset of, and thus in agreement with, the clusters obtained for $k$=5.

When considering only the previously studied flux coupling types, i.e., full, partial, and directional coupling, the three clustering algorithms give a consistent clustering for $k=14$ and $k=21$, corresponding to cutoff distances of 0.075 and 0.033, respectively. These sets of clusters, as well as the clusters for $k=5$, differ significantly from the five clusters obtained from the complete set of flux coupling profiles (adjusted Rand Index[75] of 0.14 for $k=5$, $6.1 \cdot 10^{-14}$ for $k=14$, and 0.01 for $k=21$, respectively). Moreover, we do not observe phylogenetic or functional relationships in these clusters (cf. *Results* and Supplementary Fig. 3).

To determine whether the clustering of flux coupling profiles based on the complete set of coupling types is a consequence of structural determinants, we employ the same clustering approach to the normalized cumulative singular value spectra of the metabolic networks[17]. The singular values of a stoichiometric matrix $S$ are given by the diagonal entries of $D$ obtained by singular value decomposition:

$$S = UDV^*.$$

The cumulative singular value spectra are obtained from the singular values by dividing the cumulative values of $D_{i,i}$ by the sum of singular values $\sum_i D_{i,i}$. Clustering of the cumulative singular value spectra yields only one set of clusters that is consistent across the clustering methods: *A. thaliana* and the genome-scale network of *H. sapiens* form two singleton clusters, while all remaining networks fall into the same cluster at a cut-off distance of 0.97, yielding $SI(C)=0.98$, $CH(C)=211.6$, and $DB(C)=0.07$. When comparing these clusters to those obtained from the flux coupling profiles, we obtain an adjusted Rand Index of -0.06, indicating a significant dissimilarity.

**Biochemically feasible network randomization and p-values.** To test the hypothesis that the flux coupling profiles of the analyzed metabolic networks reflect functionally important features of metabolism, we determine their statistical significance using network randomization under mass-balance constraints[14]. The reactions of a metabolic network are randomized by replacing their substrates and products by compounds from the same network and changing their stoichiometric coefficients, while preserving atomic mass balance. The resulting randomized networks satisfy basic physical principles, allowing us to estimate the significance of network properties in a biological context[15]. We calculate the flux coupling profile of each randomized network and determine z-scores

$$z_i = \frac{(x_i - \overline{y}_i)}{\sigma_i},$$

where $x_i = \phi_i(S)$ is the relative frequency of coupling type $i$ in the network $S$, $\overline{y}_i$ is the average frequency of coupling type $i$ over randomized networks, and $\sigma_i$ its standard deviation. The p-values are given by

$$p_i = 2 \int_{|z_i|}^{\infty} N(0,1).$$

We find that the frequencies of directional, full, and inhibitive couplings differ significantly from random networks in 16, 18, and 19 of the 23 real-world networks, respectively (z-score based p-values 0.05) suggesting that these couplings are a result of functional or evolutionary constraints[15]. The relative frequencies of partial couplings are not significant in any of the considered networks, while the relative frequencies of anti-couplings are significant only for the network of M. pneumoniae. The latter also contains the largest number of significant coupling types (i.e., full, directional, anti, inhibitive), followed by 15 networks with three significant coupling types (i.e., full, directional, inhibitive). The profiles of A. thaliana, T. maritima, M. barkeri, and M. acetivorans do not contain significant coupling types, although the p-values of inhibitive coupling in the latter two are close to the used significance level (0.0545 and 0.057, see Supplementary Table 1). Altogether, these findings suggest that the flux coupling profiles for most of the networks reflect functional or evolutionary features of the considered metabolic pathways[15].

**Sampling of reaction activity patterns.** Since the commonly employed methods for random sampling of steady-state flux distributions generate non-zero flux values, rather than a random distribution of active and inactive reactions, they are not suitable for our approach. Therefore, here we develop two different schemes for random sampling of steady-state feasible reaction activity patterns. In the first scheme, each import reaction is specified as inactive with probability 1/2. Next, a feasible flux distribution $v \in F$ is calculated by maximizing the fluxes of export reactions with randomly chosen coefficients. This approach corresponds to specifying a random set of available nutrients while maximizing diverse combinations of export fluxes, and aims at generating diverse and biologically meaningful reaction activity patterns. The steady-state compatible activity pattern of the sample is given by the support of the generated flux vector, $\sigma=|\text{sign}(v)|$.

The second scheme generates an initial feasible reaction activity pattern $\sigma_1$ by minimizing $d_1 \cdot (2\sigma_1 - 1)$, such that $\exists v \in F$ with $\sigma_1=|\text{sign}(v)|$, where $d_1$ is a random vector in $[-1, 1]^n$. A subsequent sample $\sigma_i$ is generated by minimizing $d_i \cdot (2\sigma_i - 1)$, where $d_i = 1 - 2\sum_{j=1}^{i-1} \sigma_j/(i - 1)$ is the sign inversed mean of all previous samples, normalized to [-1,1]. Again, the feasibility of $\sigma_i$ is achieved by satisfying the existence of $v \in F$ with $\sigma_i=|\text{sign}(v)|$. This scheme aims at sampling maximally distinct activity patterns. Again, the feasible activity pattern of the sample is given by $\sigma=|\text{sign}(v)|$. The obtained results based on 500, 1,000, or 5,000 activity patterns are virtually identical for each of the two sampling schemes.

**Flux coupling graph.** The flux coupling graph (FCG) of a metabolic network is a directed labeled graph $G=(V,E)$, where $V$ is the set of all unblocked, non-essential reactions of the network, and two vertices $i$, $j$ are connected by an edge $(i,j) \in E$, if reaction $i$ is coupled to reaction $j$. Five edge labels $L=\{\omega_{\text{full}}, \omega_{\text{partial}}, \omega_{\text{directional}}, \omega_{\text{anti}}, \omega_{\text{inhibitive}}\}$ indicate that $i$ is coupled to $j$ by the corresponding coupling type. Note that edges with labels $\omega_{\text{full}}, \omega_{\text{partial}}, \omega_{\text{anti}}$ are symmetric: if $(i,j) \in E$ and $L(i,j) \in \{\omega_{\text{full}}, \omega_{\text{partial}}, \omega_{\text{anti}}\}$, then $(j,i) \in E$ with $L(i,j)=L(j,i)$. Moreover, the edges of type $\omega_{\text{full}}$, $\omega_{\text{partial}}$, and $\omega_{\text{directional}}$ reflect the transitivity of the respective couplings: if $(i,j)$, $(j, k) \in E$, $L(i,j)=L(j,k) \in \{\omega_{\text{full}}, \omega_{\text{partial}}, \omega_{\text{directional}}\}$, then $(i,k) \in E$, $L(i,j)=L(j,k)=L(i,j)$. We show that calculation of the driver reactions in large networks is computationally inexpensive due to the transitivity relations (Supplementary Discussion). For a given reaction activity pattern, we use the flux coupling graph to generate the control graph and calculate the driver reactions.

**Control graph.** For a metabolic network with $n$ reactions, a given reaction activity pattern $\sigma=\{0,1\}^n$ specifies the active reactions $\sigma^1=\{i: \sigma_i=1\}$ and inactive reactions $\sigma^0=\{j: \sigma_j=0\}$, where $|\sigma^1 \cup \sigma^0|=n$ and $\sigma^1 \cap \sigma^0=\emptyset$. The steady-state feasibility of activity patterns, i.e., $\exists v \in F$ with $|\text{sign}(v)|=\sigma$, is guaranteed by our sampling approach (see *Sampling of reaction activity patterns*). To determine the driver reactions for $\sigma$, we first generate the control graph (CG), which contains a vertex for each reaction $i$, and a directed edge $(i \to j)$, if $i$ and $j$ are coupled, and controlling the status of reaction $i$ allows us to impose the status of reaction $j$ specified by $\sigma$. Formally, the edges in the CG are specified by the adjacency matrix $M$, where $M_{i,j}=1$, if either of the following hold:

(1) $\sigma_i=\sigma_j=1$ and $L(i,j) \in \{\omega_{\text{full}}, \omega_{\text{partial}}, \omega_{\text{directional}}\}$, or
(2) $\sigma_i=0$, $\sigma_j=1$, and $L(i,j)=\omega_{\text{anti}}$, or
(3) $\sigma_i=1$, $\sigma_j=0$, and $L(i,j) \in \{\omega_{\text{inhibitive}}\}$, or
(4) $\sigma_i=\sigma_j=0$ and $L(i,j) \in \{\omega_{\text{full}}, \omega_{\text{partial}}\}$, or

(5) $\sigma_i=\sigma_j=0$ and $L(j,i)=\omega_{\text{directional}}$,

and $M_{i,j}=0$ otherwise.

**Calculation of driver reactions.** The driver reactions are given by a smallest set of reactions, whose activities must be specified to activate all reactions in $\sigma^1$ and deactivate all reactions in $\sigma^0$. In the CG, the driver reactions $D(CG)$ correspond to a minimum out-dominating set[19], i.e., a smallest set $D\subseteq V(CG)$, such that $D\cup N^+(D)=V(CG)$, where $N^+(D)$ is the first-order out-neighborhood of $D$. We determine the driver reactions from the adjacency matrix $M$ of the control graph using the following integer linear program:

$$\min \sum_{i=1}^{n} x_i$$

s.t. $x \cdot M \geq \mathbf{1}$,

where $x \in \{0, 1\}^n$. The solution yields $x$ with a minimal number of non-zero entries and $x \cdot M \geq 1$. Thus, each reaction $i$ with $x_i=1$ is in a minimum out-dominating set of the CG, and corresponds to a driver reaction of $\sigma$.

**Transcriptional regulation of *E. coli* metabolism.** To determine the agreement between coupled reactions and their co-regulation, i.e., regulation of genes by a common transcription factor (TF), we obtain the genes associated to the enzymes in the genome-scale metabolic network of *E. coli* from the Supplementary Material of Orth, et al.[12] and KEGG[76]. This results in 925 enzymatic reactions (and, hence, 427,350 reaction pairs), of which 24 are involved in transport of protons or molecules across the inner membrane: ATP synthase, fatty acid-coenzyme A ligase (involving 10 different fatty acids), formate dehydrogenase (oxidation of either ubiquinone or menaquinone-8), hydrogenase (hydrogenation of ubiquinone, menaquinone-8, or 2-Demethylmenaquinone), NADH dehydrogenase (oxidation of ubiquinone, menaquinone-8, or 2-Demethylmenaquinol-8), glycohydrolase catalyzed cytosolic import of nicotinamide, nitrate reductase (reduction of ubiquinol in the cytosol or periplasm and reduction of 2-Demethylmenaquinol-8), and NAD(P)$^+$ transhydrogenase. Next, we obtain the transcriptional regulatory network from RegulonDB[11], and determine the TFs regulating (i.e., activating or repressing) the genes of the 925 enzymatic reactions of *E. coli*. Two reactions are co-regulated, if there is at least one TF activating or repressing any of the genes associated to both reactions (cf. Supplementary Fig. 2). The number of TFs regulating a reaction is then given by the number of unique TFs activating or repressing the genes that encode its catalyzing enzymes (cf. Fig. 4). Similar results are obtained when extending the analysis to include genes regulated by small RNAs.

**Driver reactions in cancer.** We analyze four cancer networks (i.e., breast, lung, renal, and urothelial cancer) and networks of the corresponding healthy tissues (glandular breast, pneumocytes, kidney, and urothelial cells)[69]. To determine the reactions which are associated to genes known to cause cancer, we obtain 547 *census genes* from the Catalogue of somatic mutations in cancer (COSMIC)[25] and additional 1,453 genes from high-throughput mutational screenings, whole-exome sequencing, and whole-genome sequencing of cancer samples from the Network of Cancer Genes (NCG)[27] (March 2015 database versions). Next, we map the reactions to genes through their gene-enzyme-reaction relationship (which excludes non-enzymatic reactions, such as transport and diffusion reactions), yielding a total of 2,363 reactions from the four cancer networks. Of these, 112 are associated to cancer causing genes from COSMIC (4.7%), and 614 to genes from the NCG dataset (26.0%). We identify 620 reactions as critical drivers in any of the four cancer networks (26.2%). Out of these, 45 are associated to the census genes in COSMIC (7.3%) and 191 to genes from the NCG dataset (30.8%), yielding p-values of $6.8 \cdot 10^{-4}$ (COSMIC) and $9.6 \cdot 10^{-4}$ (NCG) (hypergeometric test) for the overrepresentation of these genes.

To identify reactions which may have a key role in (de)regulation of cancer metabolism and tumor development, we focus on reactions which are critical drivers in cancer networks and redundant in the corresponding healthy tissues (*metabolic switches*). There is no reaction which is a metabolic switch for each of the four cancer types. Diacylglycerol kinase is the only metabolic switch for three cancer types (lung, urothelial, and renal). Four reactions are metabolic switches for two cancer types: S-adenosylhomocysteine hydrolase (urothelial and breast), deoxyguanosine kinase (urothelial and renal), IMP dehydrogenase (urothelial and breast), and deoxyadenosine kinase (renal and breast). Diacylglycerol kinase, S-adenosylhomocysteine hydrolase, deoxyguanosine kinase, and IMP dehydrogenase play key roles in tumor development, and there is some support for the role of deoxyadenosine kinase in cancer (see *Results* and Supplementary Table 2).

## Discussion

We propose a computational framework to study the flux control of metabolic networks, which combines the computational efficiency of biased methods with the comprehensiveness of unbiased approaches. By formalizing the requirement of biological systems to exchange matter with the environment we are able to introduce two new coupling relations: anti-coupling, representing the combined essentiality of two reactions for each possible steady-state, and inhibitive coupling, representing the competition of reactions for common substrates/products. Since the presence of anti-coupling indicates condensed rather than redundant pathways, the concept may stimulate further analyses in smaller subsystems of metabolism. The obtained complete set of coupling relations covers each possibility of (de)activating a reaction by controlling active/inactive reactions to which it is coupled. The clustering of 23 metabolic networks from diverse biological contexts by their flux coupling profiles reflects phylogenetically and functionally distinct groups, and distinguishes healthy tissues from cancer cells. This indicates that the flux coupling profiles could be used as descriptors for characterization of system-dependent metabolic requirements beyond the universal structural features of metabolic networks[77].

By using an analogy to a classical graph-theoretic problem, we determine the smallest sets of reactions that must be directly controlled to achieve a feasible reaction activity pattern at steady state. We find that critical and intermittent driver reactions are under strong transcriptional regulation in the metabolic network of *E. coli*, indicating that cellular regulation relies on driver reactions to achieve efficient control of metabolism. This serves as a proof-of-concept of our optimality-based framework. By employing metabolic networks of four cancer types and their corresponding healthy tissues, our method predicts driver reactions for controlling tumor development, many of which indeed have known causal roles in cancer. We find that reactions acting as metabolic switches for more than one cancer type are key control points of tumor development and have direct interactions with multiple cellular levels, supporting the view that tumor development is controlled at the interface of gene regulation, signaling, and metabolism. Driver reactions whose role in cancer is yet unknown may serve as starting point for identifying novel therapeutic targets. Importantly, the (bio-)chemical basis of our approach facilitates insights into the molecular mechanisms of control, such as the inhibition of methyltransferases by the modified coupling relations of SAH hydrolase in breast and urothelial cancer (cf. Fig. 5). Hence, our framework can be used to gain mechanistic insights into cellular regulation in disease and engineering of biological systems.

## Acknowledgments

We thank Alisdair Fernie, Tino Krell, Jan Lisec, and Clemens A. Schmitt for critical comments on the manuscript, and Francesco Gatto and Jens Nielsen for kindly providing the cancer networks. G.B., Z.N. and Y.-Y.L. conceived this project. G.B., Z.N., A.-L.B., and Y.-Y.L. designed the research and wrote the paper. G.B., Z.N., and A.L. performed the analyses. This work was supported by a Marie Curie Intra European Fellowship within the European Union's Seventh Framework Programme (FP7/2007–2013), ERC grant agreement number 329682, the Max Kade Foundation, and the John Templeton Foundation (award #51977). Our source code is freely available at http://scholar.harvard.edu/yyl/qfc.

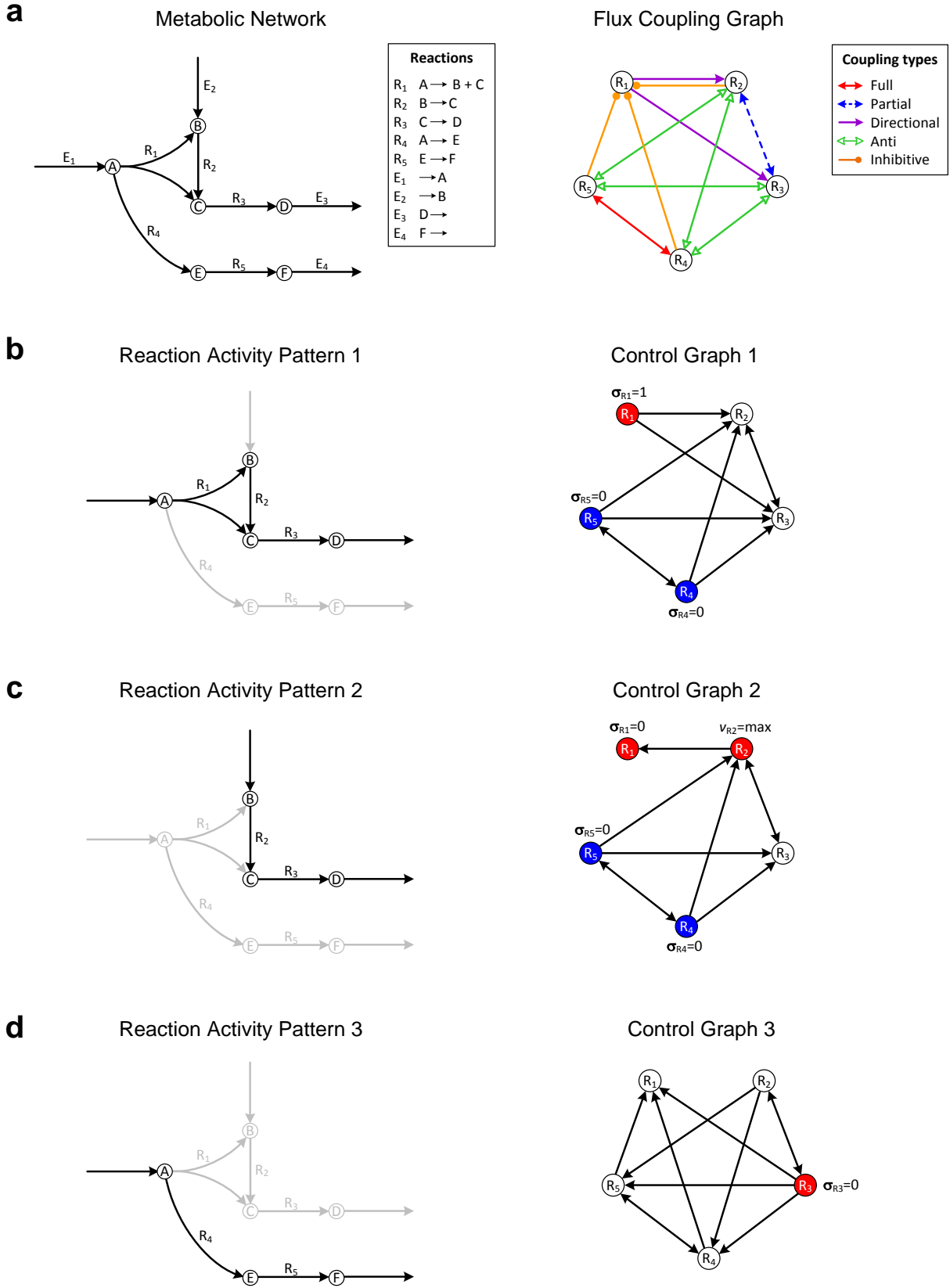

**Figure 1 | Illustration of flux coupling graph, control graph, and driver reactions.** (**a**) Left: Metabolic network with vertices representing metabolites, labeled A-F, hyper-edges representing internal reactions, labeled $R_1$-$R_5$, and exchange reactions, labeled $E_1$-$E_4$. Right: Flux coupling graph of the internal reactions of the metabolic network, with vertices

representing reactions, and labeled edges representing the five coupling relations (represented by different colors, see legend). Here, only internal reactions are considered for brevity (Supplementary Fig. 1 shows the flux coupling graph including exchange reactions). (**b**)-(**d**) Three activity patterns (left) describing the active (black arrows) and inactive reactions (gray arrows). The control graphs (right) are obtained by integrating the flux coupling graph with the activity pattern, and describe which reactions can be controlled to impose the status of other reactions in the activity pattern (see *Methods*). The driver reactions are given by a smallest set of vertices, such that each vertex in the control graph is either contained in the set or a direct successor of a vertex from the set. Vertices corresponding to driver reactions are colored; one vertex of each color must be controlled simultaneously. The required activities of the driver reactions (deactivation, activation, or maximization) are depicted next to the vertices. For example, activity pattern 1 is achieved by the simultaneous activation of $R_1$ and deactivation of either $R_4$ or $R_5$.

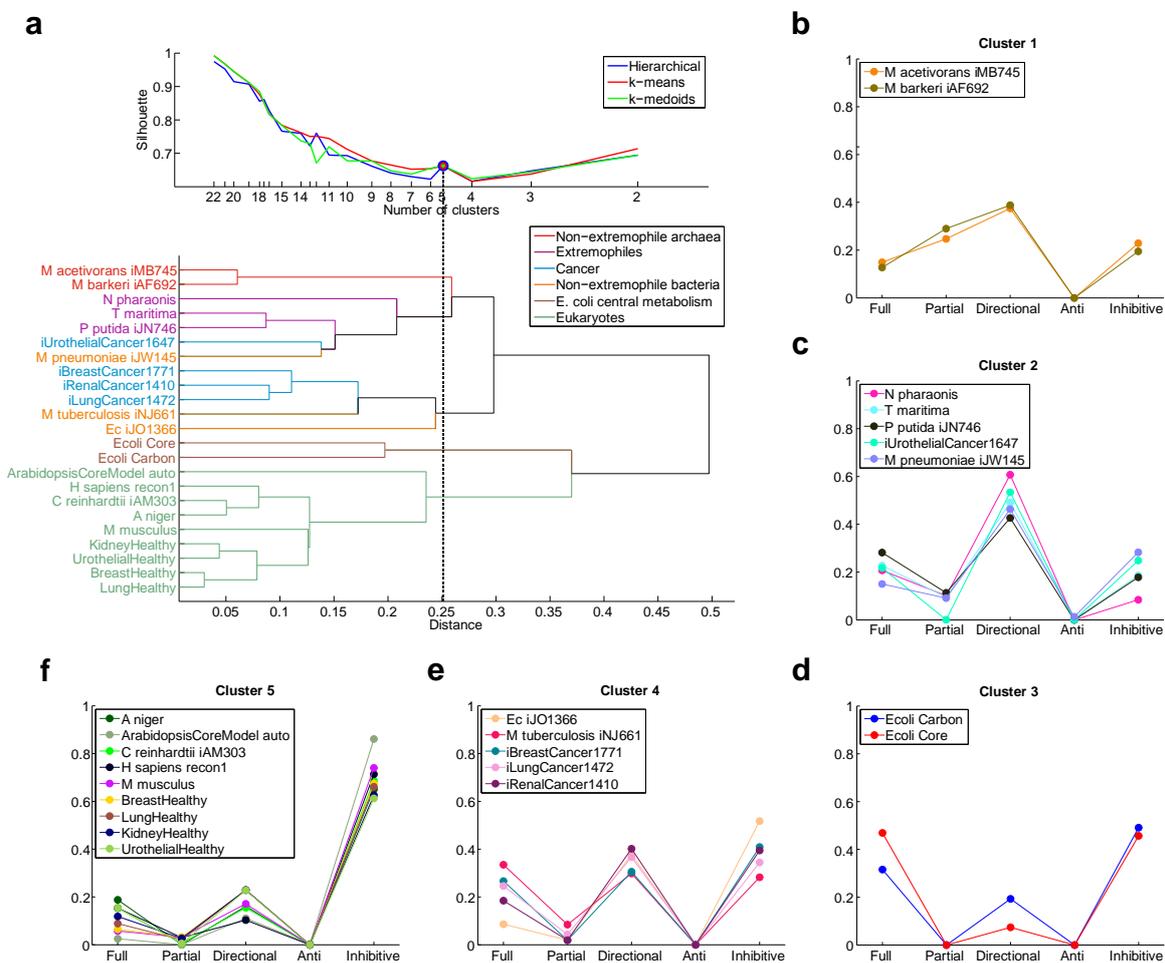

**Figure 2 | Clustering of flux coupling profiles.** (**a**) Top: Silhouette index over number of clusters for three clustering methods: hierarchical, k-means and k-medoids. For 5 clusters (dashed line) the three methods give identical clustering (blue, red, and green circles), which is shown in (**b**)-(**f**). Bottom: Cluster dendrogram obtained from hierarchical clustering of the flux coupling profiles using Euclidean distances. (**b**) Cluster 1 contains the flux coupling profiles of both non-extremophile archaea. (**c**) Cluster 2 combines the flux coupling profiles of organisms adapted to extreme environmental conditions: the haloalkaliphilic *N. pharaonis*, the hyper-thermophilic *T. maritima*, the solvent tolerant *P. putida*, as well as the pathogen *M.*

*pneumoniae* and urothelial cancer. (**d**) Cluster 3 contains the flux coupling profiles of both networks representing *E. coli* central metabolism. (**e**) Cluster 4 combines the genome-scale networks of *E. coli*, *M. tuberculosis*, and three of the four cancer networks. (**f**) Cluster 5 contains all analyzed eukaryotic networks, including the fungi *A. niger*, unicellular algae *C. reinhardtii*, *H. sapiens*, *M. musculus*, and the four tissue-specific human networks.

**a**

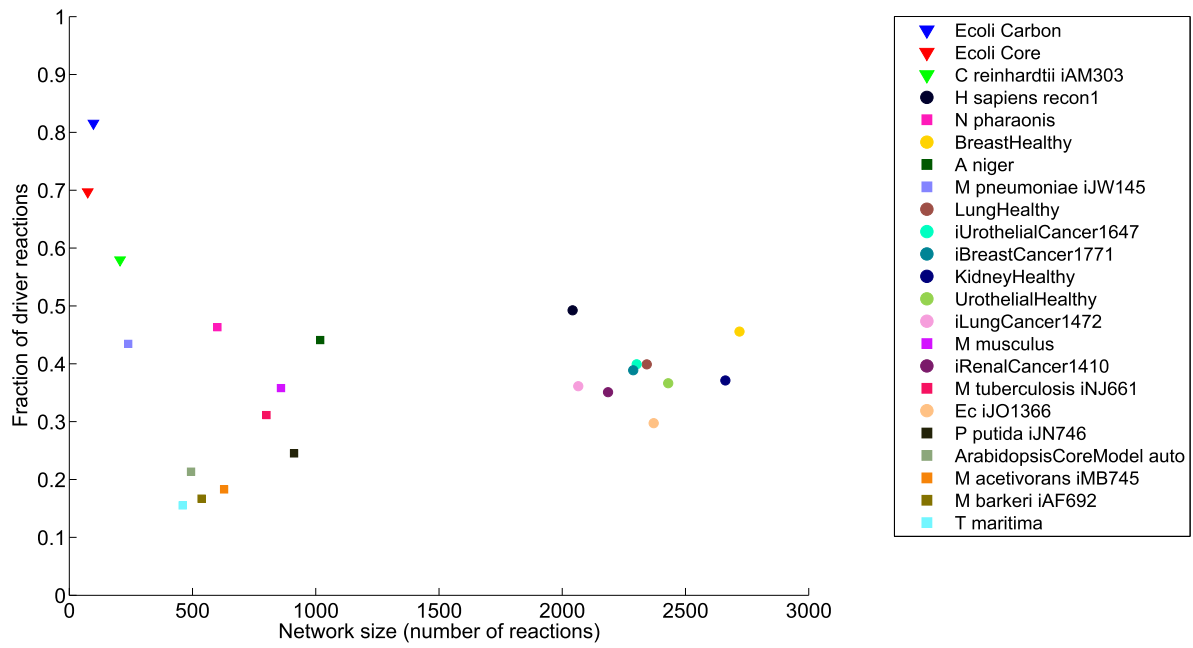

**b**

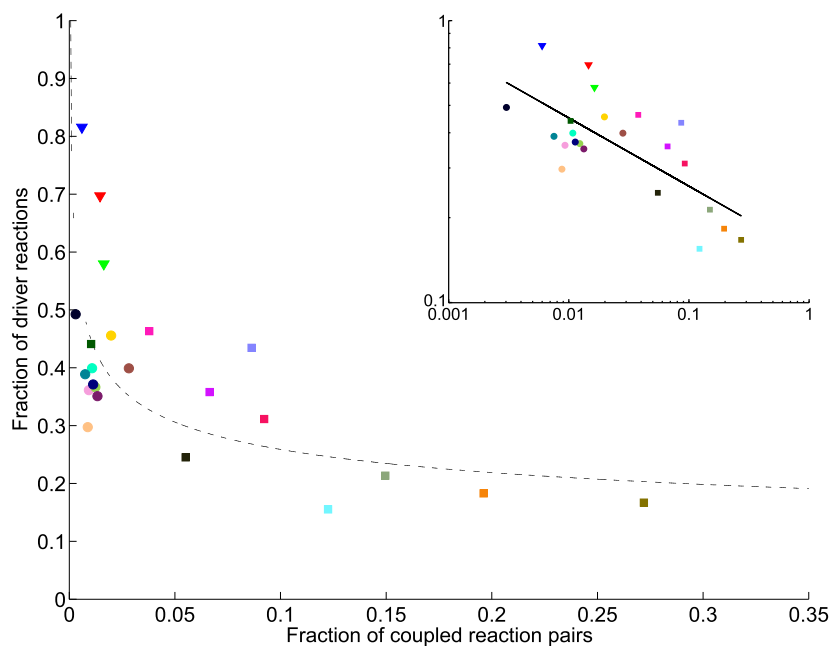

**Figure 3 | Flux control properties of the 23 analyzed metabolic networks**. (**a**) Relationship between network size (x-axis) and complexity of flux control, as quantified by the fraction of driver reactions required for controlling all reactions (y-axis). Three groups can be distinguished: complex control of small central metabolic networks (triangles), easy control of the medium sized, predominantly microbial networks (squares), and intermediately complex control of the large *E. coli*, human and cancer networks (circles). (**b**) Relationship between flux coupling and complexity of flux control. Networks arranged by the fraction of coupled reaction pairs (x-axis) and fraction of driver reactions required for controlling all reactions (y-axis). The fraction of driver reactions decays logarithmically (scaling coefficient of 0.24, dashed line) with the fraction of coupled reactions, indicated by a Pearson correlation of –0.72 on the log-log data (inset). The legend is sorted by decreasing fraction of driver reactions required for control of the network.

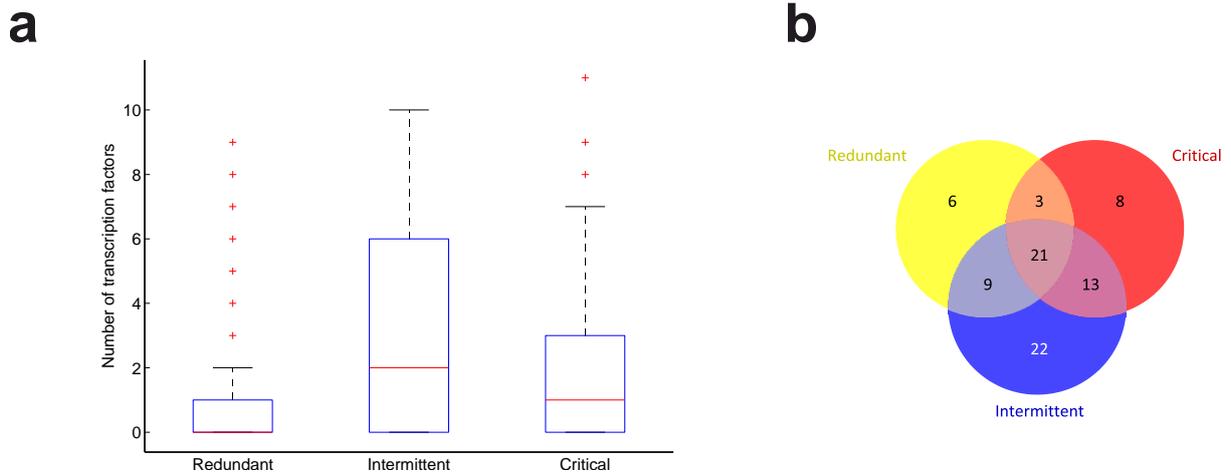

**Figure 4 | Transcriptional regulation of reaction classes in *E. coli***. (**a**) Box plot of the number of transcription factors regulating redundant, intermittent, and critical reactions, indicating the median (red lines), 25th and 75th percentiles (blue boxes), maximum non-outlier value (black whiskers), and outliers (red crosses). (**b**) Venn diagram showing the number of transcription factors associated with redundant, intermittent, and critical reactions, and their intersections.

**a** Methionine Cycle

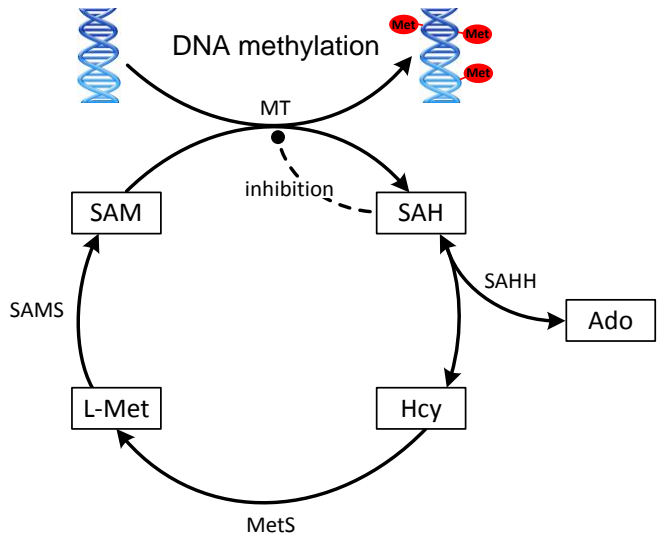

**b** Flux Coupling Graph (BreastHealthy)

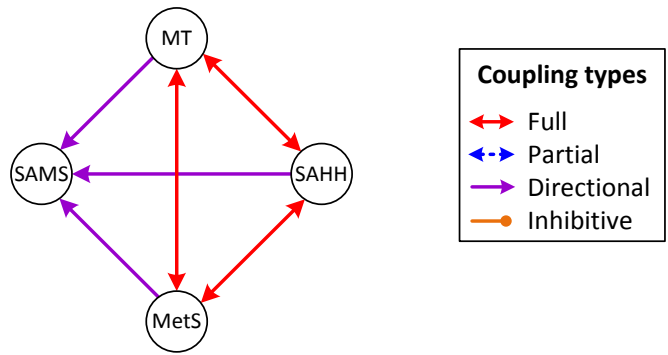

**c** Flux Coupling Graph (iBreastCancer1647)

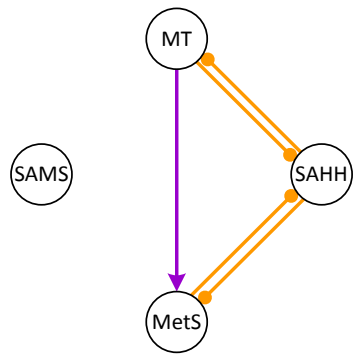

**Figure 5 | Control of methylation by the SAH hydrolase driver reaction.** (**a**) Central reactions of the methionine cycle, essential for methylation of DNA, RNA, and proteins. SAH hydrolase is identified by our method as metabolic switch for urothelial and breast cancer. It controls methylation by regulating the level of produced SAH, which inhibits methyltransferases. (**b**) Flux coupling graph of the reactions from (**a**) in healthy breast tissue. Full coupling between SAH hydrolase and methyltransferase ensures that SAH is produced and consumed at equal rates, avoiding its accumulation and DNA hypomethylation. (**c**) Flux coupling graph of the reactions from (**a**) in the breast cancer network. SAH hydrolase is no longer fully coupled to methyltransferase and methionine synthase, allowing for independent fluxes of the reactions and varying SAH levels. Inhibitive coupling between SAH hydrolase and methyltransferase implies that elevated flux of either reaction inhibits flux of the other reaction due to accumulation of the common product SAH, leading to DNA hypomethylation and cancer, which is in agreement with experimental findings (*Results*). Abbreviations: Ado, Adenosine; Hcy, Homocysteine; L-Met, L-Methionine; SAH, S-adenosylhomocysteine; SAHH, SAH hydrolase; SAM, S-adenosylmethionine; SAMS, SAM synthase; MetS, methionine synthase; MT, methyltransferase.

**Table 1 | Metabolic networks analyzed in the paper.** For each network, we show its cluster (*C*; cf. Fig. 2), network name, number of (unblocked) reactions (*n*), percent of driver reactions (*d*), brief description, and reference.

| *C* | Network | *n* | *d* | Description | Ref. |
|---|---|---|---|---|---|
| 1 | M acetivorans iMB745 | 628 | 18.3 | Genome-scale metabolic network of *Methanosarcina acetivorans* | 58 |
| | M barkeri iAF692 | 537 | 16.7 | Genome-scale metabolic network of *Methanosarcina barkeri* | 59 |
| 2 | N pharaonis | 601 | 46.3 | Genome-scale metabolic network of *Natronomonas pharaonis* | 57 |
| | T maritima | 460 | 15.5 | Core metabolism of *Thermotoga maritima* | 60 |
| | P putida iJN746 | 912 | 24.5 | Genome-scale metabolic network of *Pseudomonas putida* KT2440 | 61 |
| | iUrothelialCancer1647 | 2,302 | 39.9 | Metabolic network of urothelial cancer cells | 69 |
| | M pneumoniae iJW145 | 239 | 43.4 | Genome-scale metabolic network of *Mycoplasma pneumoniae* | 62 |
| 3 | Ecoli Carbon | 98 | 81.6 | Carbon metabolism of *Escherichia coli* | 5 |
| | Ecoli Core | 75 | 69.7 | Central metabolism of *Escherichia coli* | 68 |
| 4 | Ec iJO1366 | 2,371 | 29.7 | Genome-scale metabolic network of *Escherichia coli* | 12 |
| | M tuberculosis iNJ661 | 800 | 31.1 | Genome-scale metabolic network of *Mycobacterium tuberculosis* H37Rv | 63 |
| | iBreastCancer1771 | 2,288 | 38.9 | Metabolic network of breast cancer cells | 69 |
| | iLungCancer1472 | 2,065 | 36.1 | Metabolic network of lung cancer cells | 69 |
| | iRenalCancer1410 | 2,186 | 35.1 | Metabolic network of renal cancer cells | 69 |
| 5 | A niger | 1,018 | 44.1 | Genome-scale metabolic network of *Aspergillus niger* | 64 |
| | ArabidopsisCoreModel auto | 494 | 21.4 | Core metabolism of *Arabidopsis thaliana* under autotrophic day conditions | 65 |
| | C reinhardtii iAM303 | 206 | 57.9 | Central metabolism of *Chlamydomonas reinhardtii* | 66 |
| | H sapiens recon1 | 2,042 | 49.2 | Genome-scale metabolic network of *Homo sapiens* | 17 |
| | M musculus | 859 | 35.8 | Genome-scale metabolic network of *Mus musculus* | 67 |
| | BreastHealthy | 2,719 | 45.6 | Metabolic network of human glandular breast cells | 69 |
| | LungHealthy | 2,343 | 39.9 | Metabolic network of human pneumocytes | 69 |
| | KidneyHealthy | 2,662 | 37.1 | Metabolic network of human | 69 |

| | | | kidney cells | |
| UrothelialHealthy | 2,430 | 36.6 | Metabolic network of human urothelial cells | 69 |

# References

1. Sweetlove, L. J. & Ratcliffe, R. G. Flux-balance modeling of plant metabolism. *Front Plant Sci* **2**, 38 (2011).
2. McCloskey, D., Palsson, B. Ø. & Feist, A. M. Basic and applied uses of genome-scale metabolic network reconstructions of Escherichia coli. *Mol Syst Biol* **9**, 661 (2013).
3. Varma, A. & Palsson, B. Ø. Metabolic Flux Balancing: Basic Concepts, Scientific and Practical Use. *Bio/Technology* **12**, 994-998 (1994).
4. Lewis, N. E., Nagarajan, H. & Palsson, B. O. Constraining the metabolic genotype-phenotype relationship using a phylogeny of in silico methods. *Nat Rev Microbiol* **10**, 291-305 (2012).
5. Schuetz, R., Kuepfer, L. & Sauer, U. Systematic evaluation of objective functions for predicting intracellular fluxes in Escherichia coli. *Mol Syst Biol* **3**, 119 (2007).
6. Schuster, R. & Schuster, S. Refined algorithm and computer program for calculating all non-negative fluxes admissible in steady states of biochemical reaction systems with or without some flux rates fixed. *Comput Appl Biosci* **9**, 79-85 (1993).
7. Schilling, C. H., Letscher, D. & Palsson, B. Ø. Theory for the systemic definition of metabolic pathways and their use in interpreting metabolic function from a pathway-oriented perspective. *J Theor Biol* **203**, 229-248 (2000).
8. Terzer, M. & Stelling, J. Large-scale computation of elementary flux modes with bit pattern trees. *Bioinformatics* **24**, 2229-2235 (2008).
9. Burgard, A. P., Nikolaev, E. V., Schilling, C. H. & Maranas, C. D. Flux coupling analysis of genome-scale metabolic network reconstructions. *Genome Res* **14**, 301-312 (2004).
10. Fell, D. A. Enzymes, metabolites and fluxes. *J Exp Bot* **56**, 267-272 (2005).
11. Salgado, H. *et al.* RegulonDB v8.0: omics data sets, evolutionary conservation, regulatory phrases, cross-validated gold standards and more. *Nucleic Acids Res* **41**, D203-D213 (2013).
12. Orth, J. D. *et al.* A comprehensive genome-scale reconstruction of Escherichia coli metabolism--2011. *Mol Syst Biol* **7**, 535 (2011).
13. Notebaart, R. A., Teusink, B., Siezen, R. J. & Papp, B. Co-regulation of metabolic genes is better explained by flux coupling than by network distance. *PLoS Comput Biol* **4**, e26 (2008).
14. Basler, G., Ebenhöh, O., Selbig, J. & Nikoloski, Z. Mass-balanced randomization of metabolic networks. *Bioinformatics* **27**, 1397-1403 (2011).
15. Basler, G., Grimbs, S., Ebenhöh, O., Selbig, J. & Nikoloski, Z. Evolutionary significance of metabolic network properties. *J R Soc Interface* **9**, 1168-1176 (2012).
16. Gevers, D., Vandepoele, K., Simillon, C. & Van de Peer, Y. Gene duplication and biased functional retention of paralogs in bacterial genomes. *Trends Microbiol* **12**, 148-154 (2004).
17. Duarte, N. C. *et al.* Global reconstruction of the human metabolic network based on genomic and bibliomic data. *Proc Natl Acad Sci U S A* **104**, 1777-1782 (2007).
18. Omranian, N. *et al.* Differential metabolic and coexpression networks of plant metabolism. *Trends in Plant Science* **20**, 266 - 268 (2015).
19. Chartrand, G., Harary, F. & Yue, B. Q. On the out-domination and in-domination numbers of a digraph. *Discrete Mathematics* **197–198**, 179 - 183 (1999).
20. Garey, M. R. & Johnson, D. S. *Computers and Intractability: A Guide to the Theory of NP-Completeness*. (W. H. Freeman & Co., 1979).
21. Marashi, S.-A. & Bockmayr, A. Flux coupling analysis of metabolic networks is sensitive to missing reactions. *Biosystems* **103**, 57-66 (2011).
22. Monk, J., Nogales, J. & Palsson, B. Ø. Optimizing genome-scale network reconstructions. *Nat Biotechnol* **32**, 447-452 (2014).
23. Jia, T. *et al.* Emergence of bimodality in controlling complex networks. *Nat Commun* **4**, 2002 (2013).
24. Heiden, M. G. V., Cantley, L. C. & Thompson, C. B. Understanding the Warburg effect: the metabolic requirements of cell proliferation. *Science* **324**, 1029-1033 (2009).
25. Futreal, P. A. *et al.* A census of human cancer genes. *Nat Rev Cancer* **4**, 177-183 (2004).
26. Agren, R. *et al.* Reconstruction of genome-scale active metabolic networks for 69 human cell types and 16 cancer types using INIT. *PLoS Comput Biol* **8**, e1002518 (2012).

27  D'Antonio, M., Pendino, V., Sinha, S. & Ciccarelli, F. D. Network of Cancer Genes (NCG 3.0): integration and analysis of genetic and network properties of cancer genes. *Nucleic Acids Res* **40**, D978-D983 (2012).
28  Yi, P. *et al.* Increase in plasma homocysteine associated with parallel increases in plasma S-adenosylhomocysteine and lymphocyte DNA hypomethylation. *J Biol Chem* **275**, 29318-29323 (2000).
29  Caudill, M. A. *et al.* Intracellular S-adenosylhomocysteine concentrations predict global DNA hypomethylation in tissues of methyl-deficient cystathionine beta-synthase heterozygous mice. *J Nutr* **131**, 2811-2818 (2001).
30  Calvisi, D. F. *et al.* Altered methionine metabolism and global DNA methylation in liver cancer: relationship with genomic instability and prognosis. *Int J Cancer* **121**, 2410-2420 (2007).
31  Huang, S. Histone methyltransferases, diet nutrients and tumour suppressors. *Nat Rev Cancer* **2**, 469-476 (2002).
32  Simile, M. M. *et al.* Correlation between S-adenosyl-L-methionine content and production of c-myc, c-Ha-ras, and c-Ki-ras mRNA transcripts in the early stages of rat liver carcinogenesis. *Cancer Lett* **79**, 9-16 (1994).
33  Shrubsole, M. J. *et al.* Associations between S-adenosylmethionine, S-adenosylhomocysteine, and colorectal adenoma risk are modified by sex. *Am J Cancer Res* **5**, 458-465 (2015).
34  Sibani, S. *et al.* Studies of methionine cycle intermediates (SAM, SAH), DNA methylation and the impact of folate deficiency on tumor numbers in Min mice. *Carcinogenesis* **23**, 61-65 (2002).
35  Tan, J. *et al.* Pharmacologic disruption of Polycomb-repressive complex 2-mediated gene repression selectively induces apoptosis in cancer cells. *Genes Dev* **21**, 1050-1063 (2007).
36  Chiang, P. K. Biological effects of inhibitors of S-adenosylhomocysteine hydrolase. *Pharmacol Ther* **77**, 115-134 (1998).
37  Aarbakke, J. *et al.* Induction of HL-60 cell differentiation by 3-deaza-(+/-)-aristeromycin, an inhibitor of S-adenosylhomocysteine hydrolase. *Cancer Res* **46**, 5469-5472 (1986).
38  Borchardt, R. T., Keller, B. T. & Patel-Thombre, U. Neplanocin A. A potent inhibitor of S-adenosylhomocysteine hydrolase and of vaccinia virus multiplication in mouse L929 cells. *J Biol Chem* **259**, 4353-4358 (1984).
39  Hayden, A., Johnson, P. W. M., Packham, G. & Crabb, S. J. S-adenosylhomocysteine hydrolase inhibition by 3-deazaneplanocin A analogues induces anti-cancer effects in breast cancer cell lines and synergy with both histone deacetylase and HER2 inhibition. *Breast Cancer Res Treat* **127**, 109-119 (2011).
40  Arnér, E. S. On the phosphorylation of 2-chlorodeoxyadenosine (CdA) and its correlation with clinical response in leukemia treatment. *Leuk Lymphoma* **21**, 225-231 (1996).
41  Lotfi, K. *et al.* The pattern of deoxycytidine- and deoxyguanosine kinase activity in relation to messenger RNA expression in blood cells from untreated patients with B-cell chronic lymphocytic leukemia. *Biochem Pharmacol* **71**, 882-890 (2006).
42  Zhu, C., Johansson, M., Permert, J. & Karlsson, A. Enhanced cytotoxicity of nucleoside analogs by overexpression of mitochondrial deoxyguanosine kinase in cancer cell lines. *J Biol Chem* **273**, 14707-14711 (1998).
43  Rodriguez, J., Carlos O, Mitchell, B. S., Ayres, M., Eriksson, S. & Gandhi, V. Arabinosylguanine is phosphorylated by both cytoplasmic deoxycytidine kinase and mitochondrial deoxyguanosine kinase. *Cancer Res* **62**, 3100-3105 (2002).
44  Mérida, I., Avila-Flores, A. & Merino, E. Diacylglycerol kinases: at the hub of cell signalling. *Biochem J* **409**, 1-18 (2008).
45  Bacchiocchi, R. *et al.* Activation of alpha-diacylglycerol kinase is critical for the mitogenic properties of anaplastic lymphoma kinase. *Blood* **106**, 2175-2182 (2005).
46  Dominguez, C. L. *et al.* Diacylglycerol kinase α is a critical signaling node and novel therapeutic target in glioblastoma and other cancers. *Cancer Discov* **3**, 782-797 (2013).
47  Baldanzi, G. *et al.* Activation of diacylglycerol kinase alpha is required for VEGF-induced angiogenic signaling in vitro. *Oncogene* **23**, 4828-4838 (2004).
48  Avila-Flores, A., Santos, T., Rincón, E. & Mérida, I. Modulation of the mammalian target of rapamycin pathway by diacylglycerol kinase-produced phosphatidic acid. *J Biol Chem* **280**, 10091-10099 (2005).
49  Rambhatla, L., Ram-Mohan, S., Cheng, J. J. & Sherley, J. L. Immortal DNA strand cosegregation requires p53/IMPDH-dependent asymmetric self-renewal associated with adult stem cells. *Cancer Res* **65**, 3155-3161 (2005).
50  Jackson, R. C., Weber, G. & Morris, H. P. IMP dehydrogenase, an enzyme linked with proliferation and malignancy. *Nature* **256**, 331-333 (1975).

51. Hedstrom, L. IMP dehydrogenase: structure, mechanism, and inhibition. *Chem Rev* **109**, 2903-2928 (2009).
52. Shu, Q. & Nair, V. Inosine monophosphate dehydrogenase (IMPDH) as a target in drug discovery. *Med Res Rev* **28**, 219-232 (2008).
53. Carson, D. A. *et al.* Deoxycytidine kinase-mediated toxicity of deoxyadenosine analogs toward malignant human lymphoblasts in vitro and toward murine L1210 leukemia in vivo. *Proc Natl Acad Sci U S A* **77**, 6865-6869 (1980).
54. Chang, C. H., Brockman, R. W. & Bennett, J., LL. Purification and some properties of a deoxyribonucleoside kinase from L1210 cells. *Cancer Res* **42**, 3033-3039 (1982).
55. Manome, Y. *et al.* Viral vector transduction of the human deoxycytidine kinase cDNA sensitizes glioma cells to the cytotoxic effects of cytosine arabinoside in vitro and in vivo. *Nat Med* **2**, 567-573 (1996).
56. Ullman, B., Gudas, L. J., Cohen, A. & Martin, J., DW. Deoxyadenosine metabolism and cytotoxicity in cultured mouse T lymphoma cells: a model for immunodeficiency disease. *Cell* **14**, 365-375 (1978).
57. Gonzalez, O. *et al.* Characterization of growth and metabolism of the haloalkaliphile Natronomonas pharaonis. *PLoS Comput Biol* **6**, e1000799 (2010).
58. Satish Kumar, V., Ferry, J. G. & Maranas, C. D. Metabolic reconstruction of the archaeon methanogen Methanosarcina Acetivorans. *BMC Syst Biol* **5**, 28 (2011).
59. Feist, A. M., Scholten, J. C. M., Palsson, B. O., Brockman, F. J. & Ideker, T. Modeling methanogenesis with a genome-scale metabolic reconstruction of Methanosarcina barkeri. *Mol Syst Biol* **2**, 2006.0004 (2006).
60. Zhang, Y. *et al.* Three-dimensional structural view of the central metabolic network of Thermotoga maritima. *Science* **325**, 1544-1549 (2009).
61. Nogales, J., Palsson, B. O. & Thiele, I. A genome-scale metabolic reconstruction of Pseudomonas putida KT2440: iJN746 as a cell factory. *BMC Syst Biol* **2**, 79 (2008).
62. Wodke, J. A. H. *et al.* Dissecting the energy metabolism in Mycoplasma pneumoniae through genome-scale metabolic modeling. *Mol Syst Biol* **9**, 653 (2013).
63. Jamshidi, N. & Palsson, B. O. Investigating the metabolic capabilities of Mycobacterium tuberculosis H37Rv using the in silico strain iNJ661 and proposing alternative drug targets. *BMC Syst Biol* **1**, 26 (2007).
64. Andersen, M. R., Nielsen, M. L. & Nielsen, J. Metabolic model integration of the bibliome, genome, metabolome and reactome of Aspergillus niger. *Mol Syst Biol* **4**, 178 (2008).
65. Arnold, A. & Nikoloski, Z. Bottom-up Metabolic Reconstruction of Arabidopsis and Its Application to Determining the Metabolic Costs of Enzyme Production. *Plant Physiol* **165**, 1380-1391 (2014).
66. Manichaikul, A. *et al.* Metabolic network analysis integrated with transcript verification for sequenced genomes. *Nat Methods* **6**, 589-592 (2009).
67. Quek, L.-E. & Nielsen, L. K. On the reconstruction of the Mus musculus genome-scale metabolic network model. *Genome Inform* **21**, 89-100 (2008).
68. Orth, J. D., Fleming, R. M. T. & Palsson, B. O. Reconstruction and Use of Microbial Metabolic Networks: the Core Escherichia coli Metabolic Model as an Educational Guide. *EcoSal Plus* **28**, 245-248 (2010).
69. Gatto, F., Nookaew, I. & Nielsen, J. Chromosome 3p loss of heterozygosity is associated with a unique metabolic network in clear cell renal carcinoma. *Proc Natl Acad Sci U S A* **111**, E866-E875 (2014).
70. Larhlimi, A., David, L., Selbig, J. & Bockmayr, A. F2C2: a fast tool for the computation of flux coupling in genome-scale metabolic networks. *BMC Bioinformatics* **13**, 57 (2012).
71. Schellenberger, J., Lewis, N. E. & Palsson, B. O. Elimination of thermodynamically infeasible loops in steady-state metabolic models. *Biophys J* **100**, 544-553 (2011).
72. Mahadevan, R. & Schilling, C. H. The effects of alternate optimal solutions in constraint-based genome-scale metabolic models. *Metab Eng* **5**, 264-276 (2003).
73. Caliński, T. & Harabasz, J. A dendrite method for cluster analysis. *Communications in Statistics* **3**, 1-27 (1974).
74. Davies, D. L. & Bouldin, D. W. A cluster separation measure. *IEEE Trans Pattern Anal Mach Intell* **1**, 224-227 (1979).
75. Hubert, L. & Arabie, P. Comparing partitions. *Journal of Classification* **2**, 193-218 (1985).
76. Kanehisa, M. *et al.* Data, information, knowledge and principle: back to metabolism in KEGG. *Nucleic Acids Res* **42**, D199-D205 (2014).
77. Almaas, E., Kovács, B., Vicsek, T., Oltvai, Z. N. & Barabási, A.-L. Global organization of metabolic fluxes in the bacterium Escherichia coli. *Nature* **427**, 839-843 (2004).